\documentclass[manuscript]{aastex}
\usepackage{color}

\newcommand{\rmp}{Rev. Mod. Phys.}
\newcommand{\ap}{Adv. Phys.}
\newcommand{\pof}{Phys. Fluids}
\newcommand{\pop}{Phys. Plasmas}
\newcommand{\pss}{Planetary \& Space Sci.}
\newcommand{\sa}{Soviet Astr.}
\newcommand{\rpp}{Rev. Plasma Phys.}
\newcommand{\jfm}{J. Fluid Mech.}
\newcommand{\jpp}{J. Plasma Phys.}
\newcommand{\asr}{Adv. Space Res.}
\newcommand{\aanda}{A\& A}

\newcommand{\vB}{\mbox{\boldmath $ B $}}
\newcommand{\vb}{\mbox{\boldmath $ b $}}

\newcommand{\vF}{\mbox{\boldmath $ F $}}

\newcommand{\vU}{\mbox{\boldmath $ U $}}
\newcommand{\vu}{\mbox{\boldmath $ u $}}

\newcommand{\vvy}{\mbox{\boldmath $\hat y$}}
\newcommand{\vvz}{\mbox{\boldmath $\hat z$}}

\newcommand{\btimes}{\mbox{\boldmath $ \times $}}
\newcommand{\bnabla}{\mbox{\boldmath $ \nabla $}}

\slugcomment{To appear in Astrophys. J.}

\shorttitle{Simulations of the KH instability driven by CMEs}
\shortauthors{G\ómez et al.}

\begin{document}

\title{Simulations of the Kelvin-Helmholtz instability driven by coronal mass ejections in the 
turbulent corona}

\author{Daniel O. G\'omez\altaffilmark{1} and Edward E. DeLuca}
\affil{Harvard-Smithsonian Center for Astrophysics, 60 Garden St, Cambridge, MA 
02138}

\and

\author{Pablo D. Mininni}
\affil{Departamento de F\'\i sica, Facultad de Ciencias Exactas y Naturales, 
Universidad de Buenos Aires \& Instituto de F\'\i sica de Buenos Aires, Ciudad 
Universitaria, 1428 Buenos Aires, Argentina.}

\altaffiltext{1}{On sabbatical leave from Departamento de F\'\i sica, Facultad 
de Ciencias Exactas y Naturales, Universidad de Buenos Aires \& Instituto de 
Astronom\'\i a y F\'\i sica del Espacio, Ciudad Universitaria, 1428 Buenos 
Aires, Argentina.}

\begin{abstract}
Recent high resolution AIA/SDO images show evidence of the development of the 
Kelvin-Helmholtz instability, as coronal mass ejections (CMEs) expand in the 
ambient corona. A large-scale magnetic field mostly tangential to the interface 
is inferred, both on the CME and on the background sides. However, the magnetic 
field component along the shear flow is not strong enough to 
quench the instability. There is also observational evidence that the ambient 
corona is in a turbulent regime, and therefore the criteria for the development 
of the instability are a-priori expected to differ from the laminar case.

To study the evolution of the Kelvin-Helmholtz instability with a turbulent 
background, we perform three-dimensional simulations of the incompressible magnetohydrodynamic 
equations. The instability is driven by a velocity profile tangential to the CME-corona 
interface, which we simulate through a hyperbolic tangent profile. The turbulent background 
is generated by the application of a stationary stirring force. We compute the instability 
growth-rate for different values of the turbulence intensity, and find that the role of turbulence 
is to attenuate the growth. The fact that the Kelvin-Helmholtz instability is observed, sets an 
upper limit to the correlation length of the coronal background turbulence.

\end{abstract}


\section{Introduction}\label{intro}
Shear flows are ubiquitous in astrophysical problems, such as jet 
propagation in the interstellar medium 
\citep{ferrari1980,begelman1984,bodo1994}, the dynamics of spiral arms in 
galaxies \citep{dwarkadas1996}, cometary tails \citep{ershkovich1973,brandt1979} and  
differential rotation in accretion disks \citep{balbus1998}. It is also relevant 
in a variety of space physics problems, such as zonal flows in the atmospheres 
of rotating planets like Jupiter \citep{hasegawa1985}, the solar wind 
\citep{poedts1998} or the Earth's magnetopause \citep{parker1958}. 

Shear flows often give rise to the well-known Kelvin-Helmholtz (KH) instability 
\citep{helmholtz1868,kelvin1871}, which has been extensively studied by 
\citet{chandra1961}. It is an ideal hydrodynamic instability, that converts the 
energy of the large-scale velocity gradients into kinetic and/or magnetic energy 
at much smaller scales. The presence of a magnetic field component parallel to the shear 
flow has a stabilizing effect, and can even stall the instability if the parallel 
component of the Alfven velocity becomes larger than one half of the shear velocity jump 
\citep{lau1980,miura1982}. A similar instability condition was anticipated by 
\citet{ershkovich1973} in connection with observational evidence of KH in comet tails. 
On the other hand, an external magnetic field pointing in any direction  
perpendicular to the shear flow has no effect on the linear regime of the 
instability, and it is simply advected by the flow. 

The first observations of a KH pattern in the solar corona were reported by \citet{foullon2011}
for a 2010 November 3 event using data from the Atmospheric Imaging Assembly (AIA) on board the 
{\it Solar Dynamics Observatory} (SDO). 
\citet{ofman2011} also reported observations of a KH pattern obtained by AIA/SDO 
for an 2010 April 8 event. AIA produces high spatial resolution (pixel size of 0.6 arcsec) and 
high temporal cadence (10-20 sec) images of the Sun in several bandpasses 
covering white light, ultraviolet and extreme ultraviolet. The observed pattern 
of the Kelvin-Helmholtz instability {observed by \citet{foullon2011}} extends from about 
$70~Mm$ up to about $180~Mm$ above the solar surface ($1~Mm = 10^3~km$). When a coronal mass 
ejection (CME) expands supersonically upwards from the solar surface, a bow 
shock is formed ahead of the CME and a strong shear flow develops across the 
contact discontinuity separating the shocked ambient plasma from the ejected 
material. A similar configuration arises at the flanks of the Earth 
magnetopause, where the KH instability has also been observed and studied 
\citep{fujimoto1995,fairfield2000,nykyri2001}. More recently it was observed in connection 
to the magnetopause of other planets, such as Saturn 
\citep{masters2010} and Mercury \citep{sundberg2011}. When the supersonic solar 
wind impinges on these magnetized planets, it first crosses a bow shock (and 
becomes subsonic in the reference frame of the planet) and then circumvents 
the planet slipping through the outer part of a surface of tangential 
discontinuity known as the {\it magnetopause}, where a strong shear flow is 
generated.

The ambient corona is expected to be turbulent, as evidenced by measurements of 
non-thermal broadenings of highly ionized spectral lines. Most recent 
observations of nonthermal broadenings obtained by the Extreme-ultraviolet 
Imaging Spectrometer (EIS) on board {\it Hinode}, correspond to nonthermal 
motions in the range of $20-60~km.s^{-1}$ \citep{doschek2014}. The typical sizes 
of these nonthermal motions are sufficiently small to remain unresolved by EIS, 
whose pixel size is $2~arcsec$, and therefore its only manifestation is an 
excess in the Doppler broadening of spectral lines (i.e. beyond the thermal 
Doppler broadening).

The goal of the present paper is to study the interaction between these two 
rather dissimilar features: the large-scale laminar pattern generated by the 
ongoing KH instability, and the small-scale nonthermal motions presumably 
corresponding to a well developed turbulence. With this goal in mind, we set up 
three-dimensional simulations of the MHD equations, to study the evolution of 
the KH instability in the presence of a turbulent ambient background. 
\citet{nykyri2013} presented results from a large number of compressible 2.5D MHD simulations 
(without a turbulent background) for parameter values compatible with the observations 
of the 2010 November 3 event. This comparison is consistent with a magnetic field 
almost perpendicular to the flow plane, and therefore we make this assumption in 
our simulations. When a small-scale turbulent background is considered, the expected 
role on a large-scale flow is to 
produce the effect of an enhanced diffusivity which can be characterized through 
an effective or turbulent viscosity. The effect of this extra diffusivity on an 
ongoing instability for the large-scale flow, as it is currently the case for 
KH, is to reduce its growth rate or even to switch-off the instability 
completely. We test and basically confirm this theoretical picture with a series 
of simulations of a KH-unstable shear flow superimposed on a small-scale 
turbulent background with different turbulence intensities. The AIA observations 
showing a KH pattern are described in \S~\ref{aia} and the observed features of 
small-scale turbulence are summarized in \S~\ref{bro}. We introduce the MHD 
equations in \S~\ref{mhd} and describe the basic properties of the 
Kelvin-Helmholtz instability in \S\ref{khinst}. The characteristic features of the 
turbulent background generated in our simulations are discussed in \S~\ref{turb}
and our numerical results are shown in \S~\ref{num}. The potential consequences of 
the results presented in this paper are discussed in \S\ref{discu}, and our 
conclusions are listed in \S~\ref{conclu}.

\section{Observations}\label{obs}

\subsection{AIA observations}\label{aia}

The coronal mass ejection (CME) that occurred on 2010 November 3 near the southeast 
solar limb, showed the characteristic pattern of the KH instability on AIA 
images. This pattern has only been observed at the highest temperature channel, 
centered at the $131 \AA$ bandpass at $1.1\ 10^7\ K$. The sequence of AIA images 
shows a regular array of three to four vortex-like structures on the northern 
flank of the CME, that were interpreted by \citet{foullon2011} as the 
manifestation of an ongoing KH instability. The geometrical setup of a CME 
expanding upwards from the solar surface is similar to 
the one taking place at the Earth's magnetopause \citep{foullon2011}. In view of 
this similarity, these authors termed \textit{CME-pause} to the surface of 
tangential discontinuity that separates the plasma of the ejecta from the 
shocked plasma of the ambient corona.

From these observations, \citet{foullon2013} were able to estimate several of 
the relevant physical parameters for this instability, while the values of other 
parameters were inferred under different assumptions discussed in their 
subsection 5.3. The observational values for these various parameters are listed 
in \textit{Table 2} of \citet{foullon2013}. Among the most important parameters, 
they estimated a wavelength for the observed KH pattern of $\lambda = 18.5 \pm 
0.5~Mm$ and an instability growth rate of $\gamma_{KH} = 0.033 \pm 
0.012~s^{-1}$, which was driven by the velocity jump accross the shear layer of 
$680 \pm 92~km.s^{-1}$. These numbers are in good agreement with the dispersion 
relationship of the KH instability (see \S~\ref{khinst} below). The total magnetic field 
reported by \citet{foullon2013} at the CME-pause is sufficiently strong 
to correspond to Alfven speeds comparable to the velocity jump accross the shear 
layer. However, as noted by these authors, the field is largely tangential to the interfase 
and normal to the KH flow. As a result, this large Alfven speed does not play 
any significant role in the development of the instability. In a series of compressible 
2.5D MHD simulations \citet{nykyri2013} managed to approximately reproduce the observed features 
of this KH event (see more details in \S\ref{khinst}).

\citet{ofman2011} also reported observational evidence of the occurrence of the 
KH instability at the interface between a CME and the surrounding corona. Their  
event took place on 2010 April 8, it was the first to be detected in 
EUV in the solar corona and was clearly observed by six out of the seven wave 
bands of AIA/SDO. The velocity jump accross the shear layer for this event was 
estimated in the range of $6-20~km.s^{-1}$, while the wavelength of the observed 
KH pattern was $\lambda \simeq 7~Mm$, based on the size of the initial ripples. 
From the dispersion relationship corresponding to an incompressible fluid 
with a discontinuous velocity jump, an instability growth rate of $\gamma_{KH} 
\simeq 0.005~s^{-1}$ can be obtained. This value shows a reasonable agreement with 
the approximately $14~min$ over which the KH pattern was observed to grow and 
reach saturation \citep{ofman2011}. The KH features, however, were observed to 
last for as long as 107\ min. These observations were also compared with the 
results of compressible 2.5D MHD simulations, showing a good qualitative agreement 
during the nonlinear stage as well. Another KH event took place on 2011 February 24 
in connection with a CME. \citet{mostl2013} reported the quasi-periodic vortex 
structures observed by AIA/SDO and interpreted these observations with the aid 
of 2.5D MHD simulations. They find a reasonable agreement between the numerical 
results and the observations, assuming that the ejecta is about ten times denser 
than the surrounding ambient plasma. 

\subsection{Turbulent broadening}\label{bro}

Spectroscopic studies of coronal spectral lines show quantitative evidence of 
the existence of spatially unresolved fluid motions through the nonthermal 
broadening effect on these lines. Early observations were performed by a number 
of instruments, such as the slit spectrograph aboard Skylab \citep{mariska1992},
the High Resolution Telescope Spectrograph rocket \citep{bartoe1982}, the 
Solar Ultraviolet Measurements of Emitted Radiation (SUMER) aboard the Solar 
and Heliospheric Observatory \citep{teriaca1999}, or the various Solar Extreme Ultraviolet
Research Telescope and Spectrograph flights between 1991 and 1997 \citep{coyner2011}.

More recently, \citet{doschek2014} report nonthermal motions 
with velocities between $20$ and $60~km.s^{-1}$ obtained by EIS on 
\textit{Hinode}, corresponding to regions at the loop tops and above the loop 
tops during several flares.  EIS obtains images at the following two wavelength 
bands: $170-213~\AA$ and $250-290~\AA$. The angular resolution for the flare 
observations performed by \citet{doschek2014} is about $2~arcsec$. The 
line-of-sight motions responsible for these nonthermal broadenings correspond to 
plasma at temperatures in the range of $11-15~MK$, and they increase with the 
height above the flare loops.

These fluid motions have also been observed with EIS/Hinode in non-flaring 
active region loops \citep{doschek2008}. These fluid motions are being carried 
out by plasma at temperatures of about $1.2-1.4~MK$ with particle densities 
spanning the range of $5\ 10^8-10^{10}~cm^{-3}$. The rms values for the fluid 
velocities were in the range of $20-90~km.s^{-1}$. Outflow velocities in the 
range of $20-50~km.s^{-1}$ have also been detected through net blueshifts of the 
same spectral lines. The magnitude of the outflow velocities was found to be 
positively correlated with the rms velocity. \citet{brooks2011} performed a 
detailed study on active region AR 10978 using EIS/Hinode during a time span of 
five days in 2007 December. Persistent outflows were observed to take place at 
the edges of this active region, with an average speed of $22~km.s^{-1}$ and 
average rms velocities of $43~km.s^{-1}$. More recently, \citet{tian2012} 
studied upflows in connection to dimming regions generated by CMEs, and reported 
velocities of up to $100~km.s^{-1}$. It is speculated that these persistent 
outflows can be a significant source for the slow solar wind.

\section{Magnetohydrodynamic description}\label{mhd}
The incompressible MHD equations for a fully ionized hydrogen plasma are the 
Navier-Stokes equation and the induction equation
\begin{eqnarray}
\frac{\partial \vU}{\partial t} & = & - \left( \vU \cdot \bnabla \right) \vU + 
v_A^2 \left( \bnabla\btimes\vB\right) \btimes \vB - \bnabla P  + \nu \nabla^2 
\vU + \vF \label{eq:NS}\\
\frac{\partial\vB}{\partial t} & = & \bnabla\btimes\left( \vU\btimes\vB\right) + 
\eta \nabla^2 \vB \label{eq:ind}\ .
\end{eqnarray}
The velocity $\vU$ is expressed in units of a characteristic speed $U_0$, the 
magnetic field $\vB$ is in units of $ B_0$, and we also assume a characteristic 
length scale $L_0$ and a spatially uniform particle density $n_0$. In general terms, the 
assumption of incompressibility is valid provided that the plasma velocity 
associated with the instabilities being considered (i.e. the fluctuating part of the velocity 
profile), remains significantly 
smaller than the speed of sound. Note that it is only the inhomogeneous part of the velocity 
field the one that should be much smaller than the speed of sound. This might be a good assumption 
for some KH events, while other KH events might require to include compressible effects. Notwithstanding, 
in the present paper we adopt incompressibility as a simplifying assumption. Because 
of quasi-neutrality, 
the electron and the proton particle densities are equal, i.e., $n_e = n_i = 
n_0$. The (dimensionless) Alfven speed is then $v_A = B_0/\sqrt{4\pi m_i 
n_0}U_0$, while $\eta $ and $\nu$ are respectively the dimensionless magnetic 
diffusivity and kinematic viscosity. Note that for simplicity we assume isotropic 
expressions for both dissipative effects, even though in the presence of magnetic 
fields a tensor representation would be a more appropriate model \citep{braginskii1965}. 
These equations are complemented by the solenoidal conditions for both vector fields, i.e.,
\begin{equation}
\bnabla\cdot\vB = 0 = \bnabla\cdot\vU\ .
\label{eq:div}
\end{equation}

\section{Kelvin-Helmholtz instability}\label{khinst}
\begin{figure}
\epsscale{.80}
\plotone{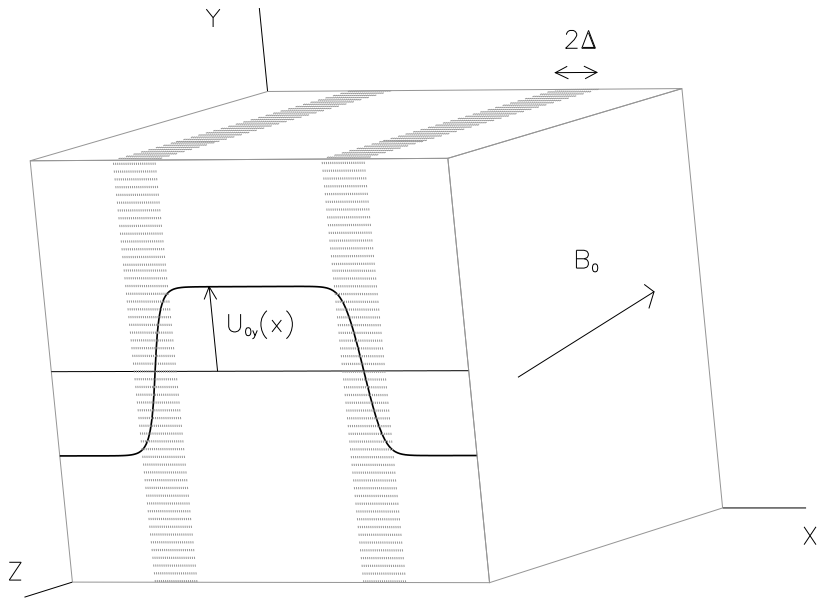}
\caption{Numerical box displaying the imposed velocity profile $U_0(x)$ in the 
${\hat y}$-direction and the external homogeneous 
magnetic field $\vB_0$. The shaded patches correspond to regions with 
intense shear. Each axis ranges from $0$ to $2\pi$.\label{fig:fig1}}
\end{figure}

Let us assume that the plasma is subjected to an externally applied shear flow 
given by
\begin{equation}
\vU_0 = U_{0y}(x) \vvy ,
 \label{eq:shear}
\end{equation}
so that the total velocity field is now $\vU_0 + \vu$, where
\begin{equation}
U_{0y}(x) = U_0 
\left[\tanh\left(\frac{x-\frac{\pi}{2}}{\Delta}\right)-\tanh\left(\frac{x-\frac{
3\pi}{2}}{\Delta}\right)-1\right]\ ,
\label{eq:tanh-shear}
\end{equation}
The velocity profile given in Eqn.~(\ref{eq:tanh-shear}) simulates the encounter 
of largely uniform flows of intensities $+ U_0 \vvy$ and $- U_0 \vvy$ through a 
parallel interface of thickness $2\Delta $. The numerical setup is 
sketched in Figure~\ref{fig:fig1}, where the jump provided by the hyperbolic 
tangent is duplicated to satisfy periodic boundary conditions throughout the 
numerical box. 
Also, we assume the presence of an external and uniform magnetic field $\vB_0$ 
tangential to the interfase and almost perpendicular to the shear flow (see 
Fig.~\ref{fig:fig1}), so that the total magnetic field is 
$\vB_0 + \vb$. The assumption of a hyperbolic tangent velocity profile is often adopted 
\citep{drazin1958,chandra1961,miura1992} as a way to model shear flows with a 
finite thickness. The velocity profile given in Eqn.~(\ref{eq:tanh-shear}) is an 
exact equilibrium of Eqs.~(\ref{eq:NS})-(\ref{eq:ind}) obtained through the 
application of the stationary external force $\vF_0 = -\nu\nabla^2 
U_{0y}(x)\vvy$ (see also \citet{gomez2014}), and therefore it is numerically 
implemented simply by the application of the volume force $\vF_0$.  

In the KH event that took place at one of the flanks of the 2010 November 3 CME, 
the fluid is observed to move along the contact discontinuity, albeit at very different speeds on 
either side. We choose to describe the development of the KH instability from a reference 
frame moving along the interfase at the average between these two speeds. In this reference 
frame, the flow will display a hyperbolic tangent type of profile, for which the parameter $U_0$ 
(see Eqn~(\ref{eq:tanh-shear}))  will be equal to one half of the relative velocity.

A shear flow such as the one given by Eqn.~(\ref{eq:tanh-shear}) is subjected to 
the well known Kelvin-Helmholtz instability, 
which is of a purely hydrodynamic nature, i.e. it occurs even in the absence of 
any magnetic field. Within the framework of MHD, the 
stability of a tangential velocity discontinuity (i.e. in the limit of $\Delta 
=0$) was first studied by \citet{chandra1961}. For the case of an external 
magnetic field aligned with the shear flow, the mode is stabilized by the 
magnetic field, unless the velocity jump exceeds twice the Alfv\'en speed. For 
the case at hand, we assume the parallel component of the external magnetic 
field to be sufficiently weak (i.e. $v_A^\parallel < 1$), since otherwise the 
instability pattern would not have been observed in AIA images. A stability analysis 
of a sheared MHD flow of finite thickness (i.e., $\Delta \ne 0$) in a compressible plasma 
has also been performed \citep{miura1982}, confirming the result of the purely hydrodynamic case. 
Compressibility has a stabilizing effect in the sense that the growth rate is reduced as the 
velocity jump approaches the speed of sound, and even stalls the instability when the Mach 
number becomes unity \citep{miura1982}. From {\it Table 2} of \citet{foullon2013}, we derive a shear 
flow amplitude $U_0 = 340 km.s^{-1}$, which remains below the speed 
of sound at both sides of the CME-pause. For the sake of simplicity, we neglect the 
effect of compressibility, which would bring an extra parameter to the problem: the Mach number. Yet 
another effect that might become relevant for the evolution of the KH instability, is the presence of 
a density contrast between both sides of the shear flow \citep{prialnik1986,gonzalez1994,wyper2013}. 
However, for the particular event under consideration it is not expected to play a role, since the mass 
density at both sides of the CME-pause remains virtually the same \citep{foullon2013}. 

\begin{figure}
\epsscale{.80}
\plotone{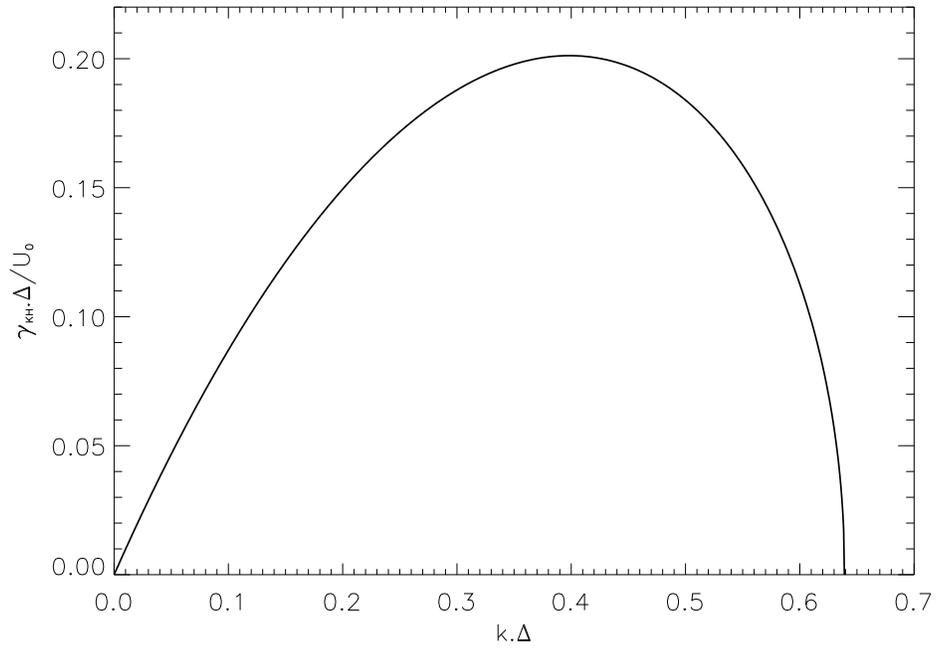}
\caption{Growth rate for the Kelvin-Helmholtz instability of a shear layer with 
a velocity jump from $+U_0$ to $-U_0$ over a 
half-width $\Delta$ as a function of wavenumber.\label{fig:fig2}}
\end{figure}

If we approximate the hyperbolic tangent profile given in Eqn.~(\ref{eq:tanh-shear}) 
by piecewise linear functions, the instability growth rate $\gamma_{KH}$ arising 
from the linearized version of Eqs.~(\ref{eq:NS})-(\ref{eq:ind}) is 
(for details, see \citet{drazin1981})
\begin{equation}
\left(\frac{\gamma_{KH}\Delta}{U_0}\right)^2 = \frac{1}{4}\left(e^{-4k_y\Delta} 
- (2k_y\Delta - 1)^2\right) ,
\label{eq:gamma}
\end{equation}
which attains its maximum at $\lambda_{max}\approx 15.7\ \Delta$ and 
$\gamma_{max}\approx 0.2 U_0/\Delta$, as shown in Figure~\ref{fig:fig2}. More 
importantly, Figure~\ref{fig:fig2} also shows that the KH instability only 
arises for large-scale modes, i.e. such that $k_y\Delta\le 0.64$, corresponding 
to $\lambda \ge 9.82\Delta$.

We perform numerical integrations of Eqs.~(\ref{eq:NS})-(\ref{eq:ind}) subjected 
to the external force $\vF_0 = -\nu\nabla^2 U_{0y}(x)\vvy$ (where $U_{0y}(x)$ is 
given in Eqn.~(\ref{eq:tanh-shear})) on the cubic box of linear size $2\pi$ 
sketched in Figure~\ref{fig:fig1}, assuming periodic boundary conditions in all 
three directions. The number of gridpoints is $256^3$ and the dimensionless 
Alfven speed was set at $v_A^\parallel = 0.2$ in all our simulations, indicating that the 
component of the external magnetic field parallel to the flow (i.e. $B_{0y}$, see 
Fig.~\ref{fig:fig1}) is such that its associated Alfven velocity component remains 
smaller than the maximum velocity $U_0$ of the shear profile, and it is therefore unable 
to quench the instability. This is indeed the case of the 2010 November 3 KH event. 
\citet{nykyri2013} performed a series of 2.5D MHD simulations seeking to match the time 
development of the KH pattern observed by AIA/SDO. Their numerical quest is consistent 
with slightly different magnetic field intensities at either side of the shear layer within 
the range of 8-11 G, forming small angles with the $\hat{z}$-direction (between $1^\circ$ 
and $10^\circ$, see Fig.~\ref{fig:fig1}), which leads to values of $v_A^\parallel$ in the 
range of $v_A^\parallel \approx 0.04-0.31$.

In our simulations, we use a 
pseudospectral method to perform the spatial derivatives and a second order 
Runge-Kutta scheme for the time integration (see a detailed description of the 
code in \cite{gomez2005}). For the viscosity and resistivity coefficients we 
chose $\nu = \eta = 2.10^{-3}$, which are small enough to produce energy 
dissipation only at very small scales, comparable to the Nyquist wavenumber. In 
particular, dissipative effects are certainly negligible for all wavenumbers becoming 
unstable due to KH (see Eqn.~(\ref{eq:gamma}) and the text right below it). The 
values of all the dimensionless parameters adopted for our simulations are 
summarized in Table~\ref{table}. In all simulations, the pressure in 
Eqn.~(\ref{eq:NS}) is obtained self-consistently by taking the divergence of the 
equation, using the incompressibility condition, and solving at each time step 
the resulting Poisson equation for the pressure.

\begin{figure}
\epsscale{.80}
\includegraphics{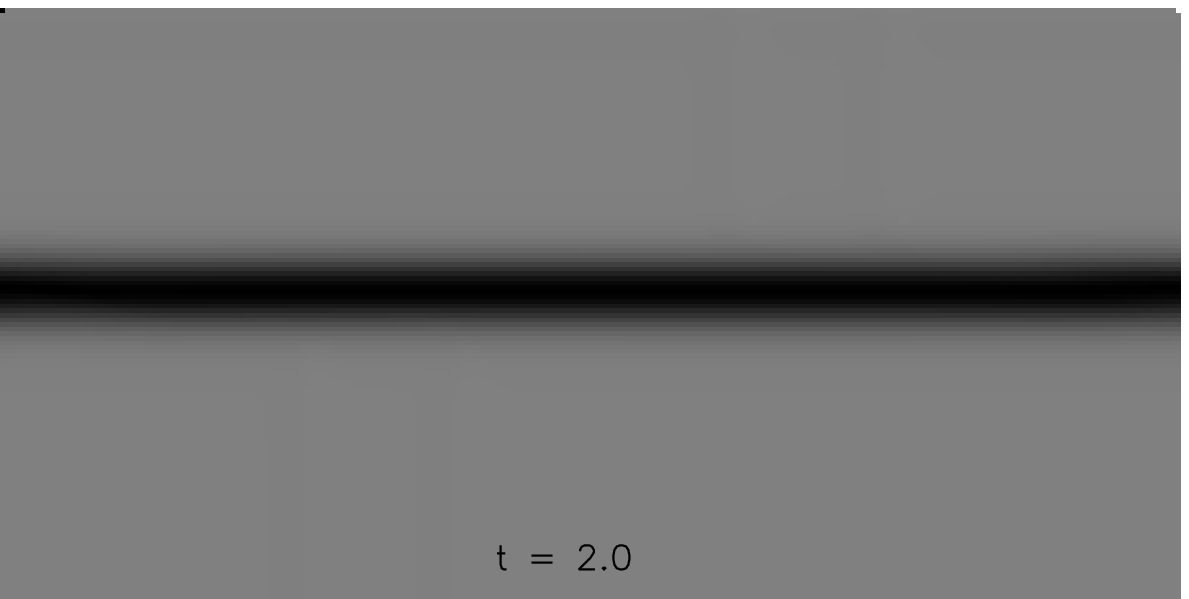}\vskip.1truecm\includegraphics{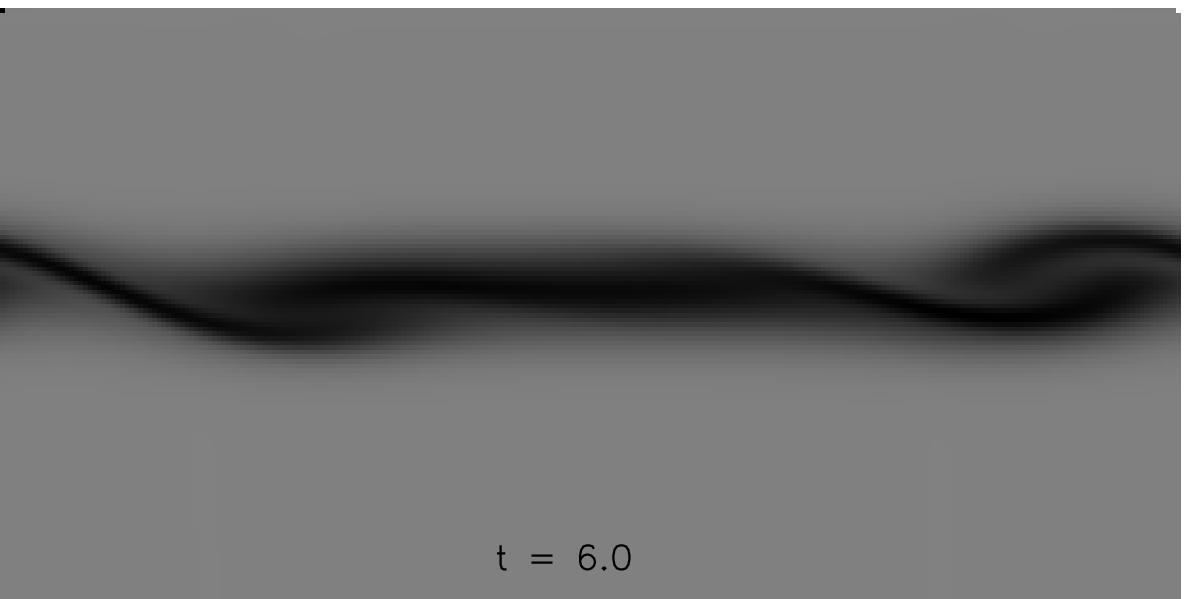}
\vskip.1truecm\includegraphics{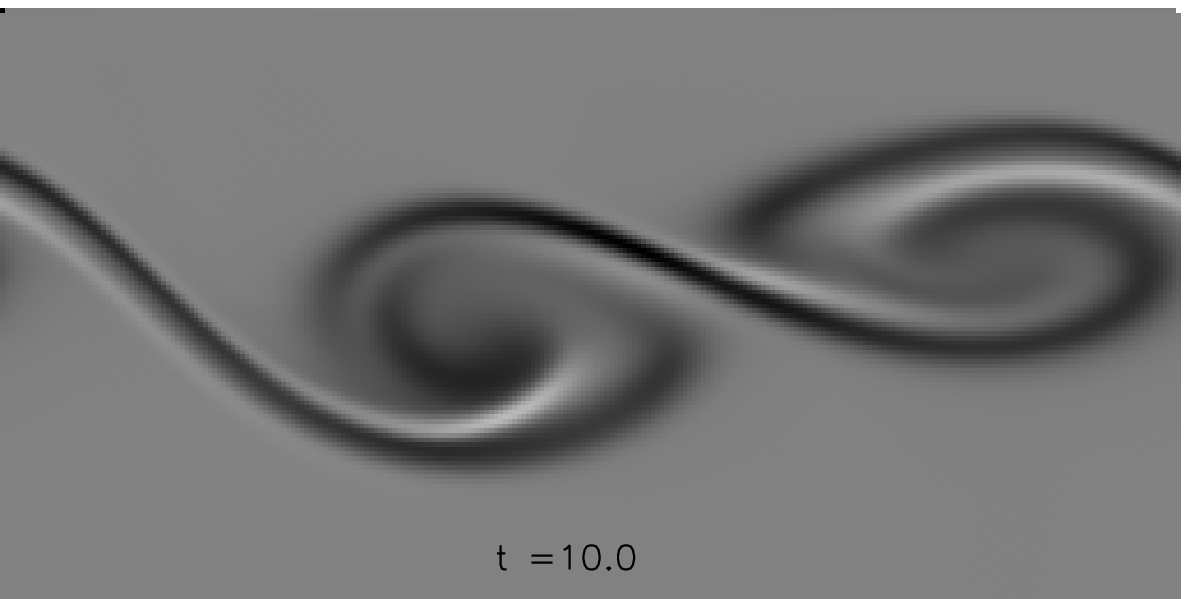}
\caption{Time sequence (as labelled) of the vorticity component $\omega_z(x,y)$ 
at the plane $z=2\pi$ for the right half of the numerical 
box shown in Fig.~\ref{fig:fig1} (rotated $90^{\circ}$) for a purely 
shear-driven simulation. Gray corresponds to $\omega_z=0$ while black 
(white) corresponds to negative (positive) concentrations of 
vorticity.\label{fig:fig3}}
\end{figure}

The evolution of the $\vvz$-component of vorticity is shown in 
Figure~\ref{fig:fig3} at three different times, displaying the characteristic 
pattern of the KH instability. The observed frame corresponds to the right half 
of the numerical box displayed in Figure~\ref{fig:fig1}, which covers the shear 
layer centered at $x_0=3\pi/2$, and has been rotated for better viewing. The 
observed pattern shows the presence of the largest Fourier mode in our numerical 
box, characterized by $k_y=1$, whose growth rate according to Eqn.~(\ref{eq:gamma}) is 
$\gamma_{KH}(k_y = 1)\simeq 0.87$. At the same time, the presence of harmonics is also apparent, 
judging by the smaller scale patterns showing up as the instability progresses. In fact, from Eqn.~(\ref{eq:gamma}) (see also Figure~\ref{fig:fig2}) we can anticipate which ones would 
be the growing Fourier modes.

\begin{figure}
\epsscale{.80}
\plotone{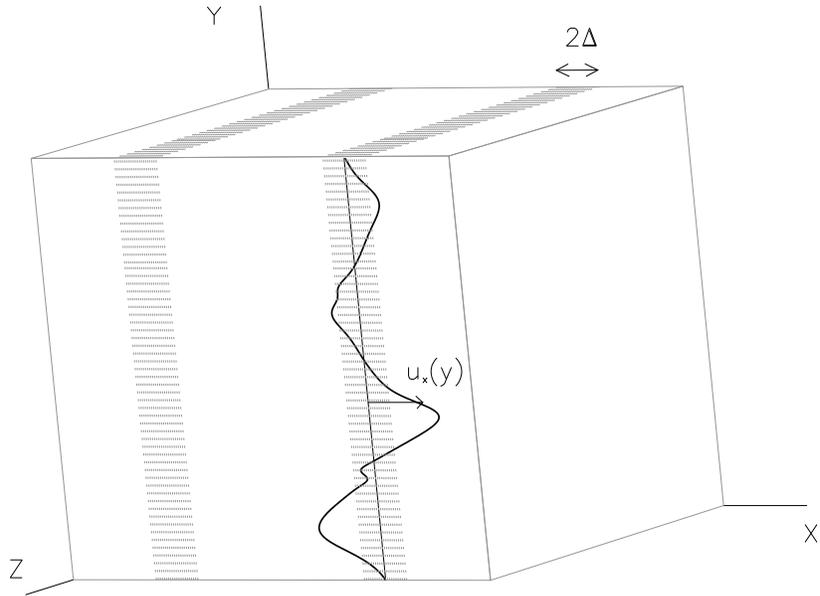}
\caption{Numerical box (see also Fig. \ref{fig:fig1}) displaying the velocity 
profile $u_x(y)$ for the slice located at the center of the shear layer. This 
velocity profile obtained for a sequence of times is used to estimate the 
instability growth rate.\label{fig:fig4}}
\end{figure}

To estimate the instability growth rate, we use the component $u_x(x_0,y,z)$ 
evaluated at $x_0=\pi/2, 3\pi/2$ (i.e., in the central part of the shear flows) 
as a proxy (see Figure~\ref{fig:fig4}). A Fourier analysis performed on $u_x(x_0,y,z)$ 
for any fixed value $z$, confirms that the exponentially growing modes belong to the 
interval $k_y=1,\dots , 6$; which is consistent with the theoretical prediction shown 
in Figure~\ref{fig:fig2} for $\Delta = 0.1$. Since the KH pattern is a two-dimensional 
flow taking place at $z=constant$ planes, we take the maximum velocity of the 
profile $u_x(x_0,y,z)$ at any given value of $z$, and then average in the 
$\vvz$-direction, i.e.
\begin{equation}\label{eq:uxmax}
U_{x,max}=\int_0^{2\pi} \frac{dz}{2\pi} \max \left[ u_x(x_0,y,z), 0\le y < 2\pi 
\right] \ . 
\end{equation}

\begin{figure}
\epsscale{.80}
\plotone{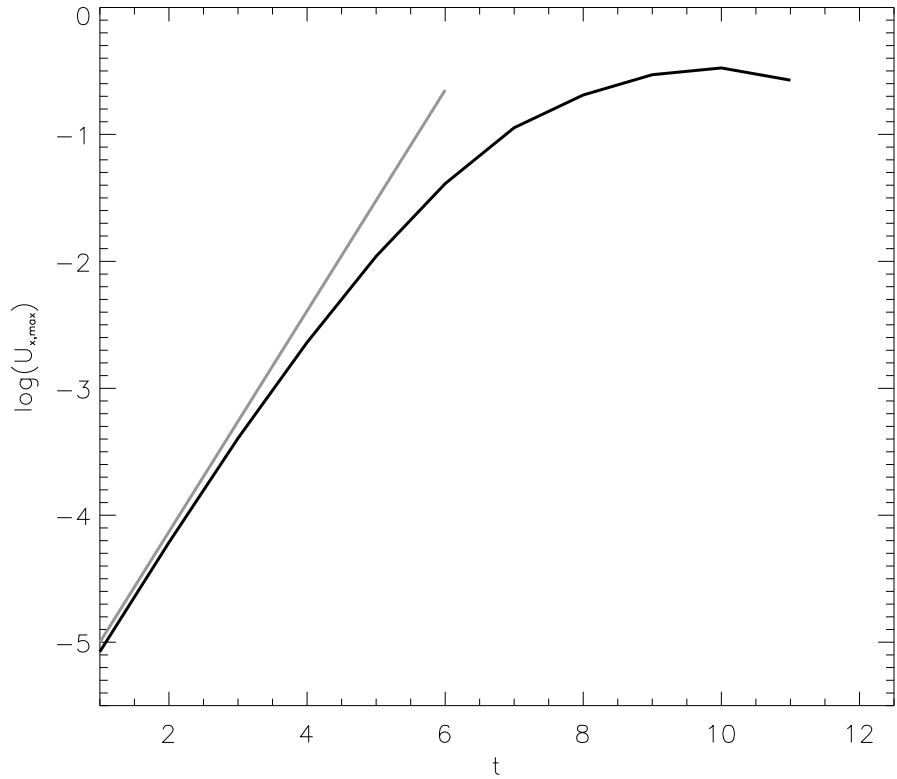}
\caption{Maximum value of the profile $u_x(x_0,y)$ vs. time in a lin-log plot. 
The two black traces are indistinguishable from one another and correspond to 
$x_0=\pi/2$ and $x_0=3\pi/2$. The straight gray line corresponds to the 
theoretical growth rate.\label{fig:fig5}}
\end{figure}

In Figure~\ref{fig:fig5} we show the maximum value of the $u_x(x_0,y,z)$ profile 
(averaged with respect to the $\vvz$-direction) for both $x_0=\pi/2$ and 
$x_0=3\pi/2$, although as expected the two curves are undistinguishable. The 
straight gray line corresponds to the theoretically predicted growth rate 
$\gamma_{KH}\simeq 0.87$ for the Fourier mode $k_y=1$ (using 
Eqn.~(\ref{eq:gamma})), which is the one observed in the time sequence 
shown in Fig.~\ref{fig:fig3}. The fact that our empirical determination of the 
growth rate so strongly resembles $\gamma_{KH} (k_y=1)$ even though (as mentioned) the 
observed pattern is more complex that a single Fourier mode, arises as the combined 
result of the $z$-averaging and our choice of the maximum of the velocity profile, 
as defined in Eqn.~(\ref{eq:uxmax}). Note that even though the simulations include 
dissipative effects and the theoretical prediction does not, the coincidence 
between both curves during the linear regime of the instability is nonetheless remarkable. 
Considering that the attenuation effect of viscosity can be estimated by $\gamma \simeq 
\gamma_{KH}-\nu k_y^2$, we can easily verify that the dissipative correction is absolutely 
negligible for the evolution of the KH instability, as expected.

\section{The turbulent corona}\label{turb}

To generate a turbulent background in our simulations, we apply a stationary 
force to all modes within a thin spherical shell of radius $k_{turb}=1/l_{turb}$ 
in Fourier space, consisting of a superposition of harmonic modes with random 
phases. The nonlinear interactions between these Fourier modes that are being 
externally driven with a force of intensity $f_{turb}$, will develop a 
stationary turbulent regime with its associated energy cascade involving all 
wavenumbers $k\ge k_{turb}$. To make sure that it is a small-scale turbulence, 
we chose $l_{turb}$ to be much smaller than the wavelength observed for the KH 
pattern, and even somewhat smaller than the thickness $\Delta$ of the shear 
layer (i.e. $l_{turb} < \Delta$). 

\begin{figure}
\epsscale{.80}
\plotone{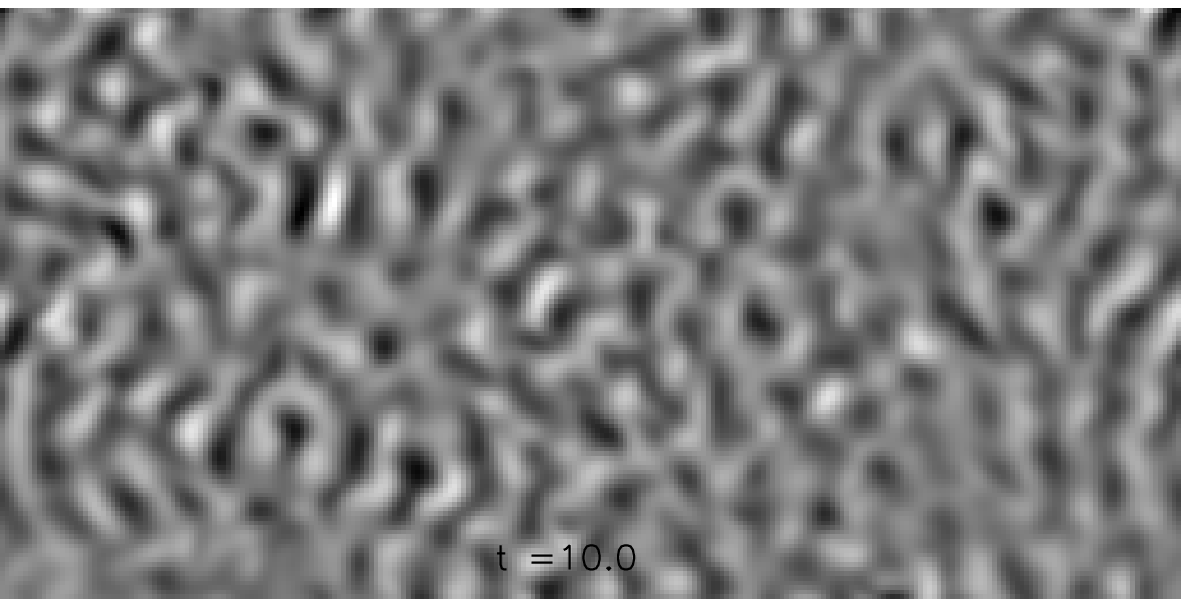}
\caption{Vorticity component $\omega_z(x,y)$ at the plane $z=2\pi$ for the right 
half of the numerical box shown in Fig. \ref{fig:fig1} (rotated $90^{\circ}$) 
for a purely turbulence-driven simulation at $t=10$. Gray corresponds to 
$\omega_z=0$ while black (white) corresponds to negative (positive) 
concentrations of vorticity.\label{fig:fig6}}
\end{figure}

\begin{figure}
\epsscale{.80}
\plotone{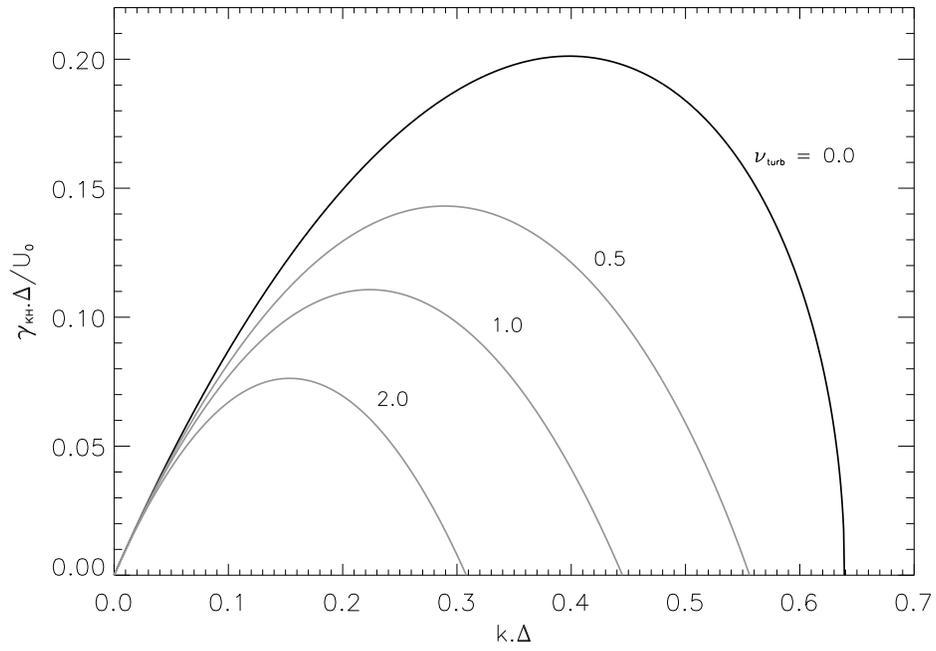}
\caption{Instability growth rates vs. wavenumber. Black trace corresponds to 
Kelvin-Helmholtz in a non-turbulent medium, as 
shown in Fig. \ref{fig:fig2}. Gray traces correspond to cases with different 
values of the turbulent viscosity $\nu_{turb}$ (labelled).\label{fig:fig7}}
\end{figure}

The pattern of vorticity obtained when only the turbulent forcing is applied 
(i.e. a simulation with no KH driving), is shown in Figure~\ref{fig:fig6}. The 
observed pattern corresponds to a turbulent regime which is statistically 
stationary, homogeneous and isotropic. Even though all spatial scales from 
$l_{turb}$ down to the smallest scales available in the simulation participate 
in the dynamics and in the ensuing energy cascade, only those vortices of sizes 
comparable to $l_{turb}$ can
be identified, which is to be expected for a power 
law power spectrum with a negative index such as Kolmogorov's. Therefore, these 
concentrations of vorticity can safely be associated to the energy-containing 
eddies of the turbulence. As mentioned in \S~\ref{intro}, the expected effect of 
this small-scale turbulence on a larger scale flow, is an effective or enhanced 
diffusivity. In the case at hand, its effect on the instability growth rate is 
expected to be 
\begin{equation}\label{eq:gamma-turb}
 \gamma(k) = \gamma_{KH}(k) - \nu_{turb} k^2\ ,
\end{equation}
where $\gamma_{KH}(k)$ is given in Eqn.~(\ref{eq:gamma}) and $\nu_{turb}$ is the 
aforementioned effective or turbulent viscosity. The effect of increasing 
turbulent viscosity on the instability growth rate is illustrated in Figure~\ref{fig:fig7}, 
showing that not only the growth rate is reduced but also the range of unstable 
wavenumbers.

\begin{figure}
\epsscale{.80}
\plotone{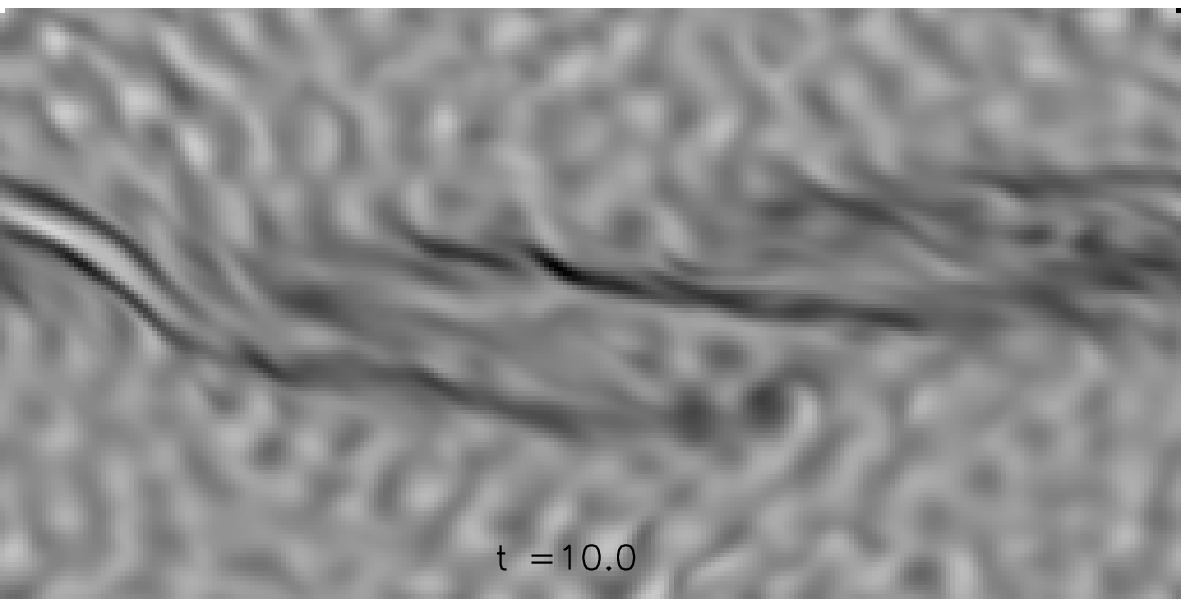}
\caption{Vorticity component $\omega_z(x,y)$ at the plane $z=2\pi$ for the right 
half of the numerical box shown in Fig. \ref{fig:fig1} (rotated $90^{\circ}$) 
for a shear and turbulence-driven simulation at $t=10$. Gray corresponds to 
$\omega_z=0$ while black (white) corresponds to negative (positive) 
concentrations of vorticity.\label{fig:fig8}}
\end{figure}

We performed simulations 
applying both the large-scale force $\vF_0$ to drive the KH instability and the 
small-scale force of intensity $f_{turb}$ to drive the turbulent regime. In 
Figure~\ref{fig:fig8} we show the resulting distribution of the vorticity 
component $\omega_z(x,y)$, which can be compared with the one shown in 
Figure~\ref{fig:fig3} for the KH instability on a laminar background, and the 
one shown in Figure~\ref{fig:fig6} for the purely turbulent run, with no KH 
pattern. We can qualitatively see that the role of turbulence is in fact an 
attenuation in the growth of the instability.

One of the observable consequences of this turbulent regime is the nonthermal 
broadening of coronal spectral lines caused by the turbulent motion of the fluid 
elements emitting these (optically thin) spectral lines. Once this turbulence 
reaches a Kolmogorov stationary regime, the rms value of the turbulent velocity 
$u_{turb}$ is
\begin{equation}\label{eq:Eturb}
 E_{turb} = \frac{u_{turb}^2}{2}=\int_{1/l_{turb}}dk\ \epsilon^{2/3}\ k^{-5/3} 
\propto (\epsilon\ l_{turb})^{2/3}\ ,
\end{equation}
where $E_{turb}$ is the (dimensionless) kinetic energy density of the turbulence 
and $\epsilon$ is its energy dissipation rate.
Note that neither $\epsilon$ or $E_{turb}$ are known a priori, since they arise 
as a result of the stationary regime attained by the 
turbulence. However, using heuristic arguments we can find how these quantities 
scale with the input parameters of this turbulence: namely $l_{turb}$ and 
$f_{turb}$. The fluid is energized by the work done per unit time by the 
external force of intensity $f_{turb}$ at scale $l_{turb}$, energy then cascades 
down to smaller scales and it is dissipated by viscosity at the rate $\epsilon$ 
at dissipative scales. In a stationary regime, the power delivered by the 
external force should match the energy dissipation rate, i.e.
\begin{equation}\label{eq:eps}
\epsilon \propto f_{turb}\ u_{turb}\ .
\end{equation}
Equations~(\ref{eq:Eturb})-(\ref{eq:eps}) can be combined to obtain both 
$u_{turb}$ and $\epsilon$ in terms of $f_{turb}$ and $l_{turb}$,
\begin{equation}\label{eq:eps-f}
\epsilon \propto ( f_{turb}^3\ l_{turb} )^{1/2}\ ,
\end{equation}
\begin{equation}\label{eq:uturb-f}
u_{turb} \propto (f_{turb}\ l_{turb})^{1/2}\ .
\end{equation}
On dimensional arguments, the turbulent viscosity introduced in 
Eqn.~(\ref{eq:gamma-turb}) has to be proportional to the turbulent velocity 
$u_{turb}$ times the typical scale $l_{turb}$, i.e. $\nu_{turb}\propto u_{turb}\ 
l_{turb}$, which considering Eqn.~(\ref{eq:uturb-f})
\begin{equation}\label{eq:nuturb}
 \nu_{turb} = C ( f_{turb}\ l_{turb}^3 )^{1/2}\ .
\end{equation}

\section{Numerical results}\label{num}

To quantify the role of turbulence in the evolution of the KH instability, we 
performed a sequence of simulations for which the only parameter being changed 
is the turbulent forcing $f_{turb}$. As the parameter $f_{turb}$ is gradually 
increased, the corresponding turbulent velocity $u_{turb}$ (observationally 
perceived as nonthermal broadening of spectral lines) is also increased, which 
in turn raises the turbulent viscosity $\nu_{turb}$. As a result, the 
instability growth rate (see Eqn.~(\ref{eq:gamma-turb})) is expected to be 
reduced. To estimate the instability growth rate from our simulations, we follow 
the same procedure described in \S~\ref{khinst}, which amounts to follow the 
temporal evolution of the profile $u_x(y)$ for the gridpoints centered at the 
shear layer. Note however, that 
now the velocity at each grid point can be split into a part corresponding to 
the large-scale KH evolution plus another part corresponding to the turbulence. 

Because of the geometrical setup of our simulations, the large-scale part of the 
flow at each $z=constant$ plane is an exact replica of one another (KH is a 
two-dimensional flow) while the turbulent part is not, since it is a fully 
three-dimensional flow. 
The averaging procedure in the $\vvz$-direction described in 
Eqn.~(\ref{eq:uxmax}) gets rid of the turbulent part of the flow, since the mean 
velocity of this turbulence is exactly zero. We can also compute the rms 
deviation of the velocity when averaging in the $\vvz$-direction, which should 
exactly correspond to $u_{turb}$, since the KH part of the flow is identical for 
all $z=constant$ planes. Therefore, this statistical strategy allows us to 
obtain the main features of both the large-scale (i.e. the KH instability) and 
small-scale (the turbulence) components of this complex flow.

\begin{figure}
\epsscale{.80}
\plotone{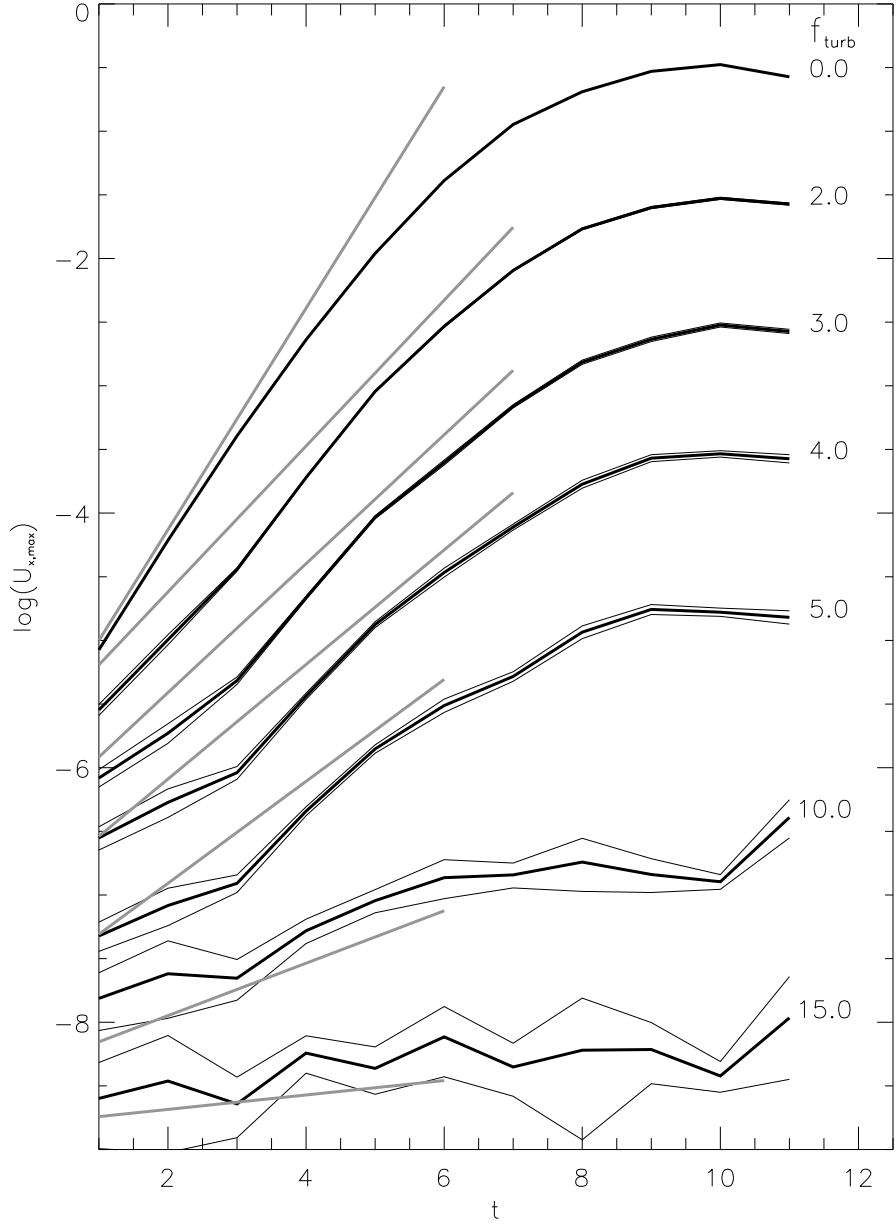}
\caption{Maximum value of the profile $u_x(x_0,y)$ vs. time in a lin-log plot 
for runs of different turbulence intensities $f_{turb}$ (labelled) and 
$x_0=\pi/2$. Each thick black trace corresponds to the average in the 
$\vvz$-direction, while the thin black traces (only noticeable for $f_{turb} = 
4$ and larger) correspond to plus or minus the root-mean deviation of the 
average. The straight gray lines correspond to the theoretical growth rate shown 
in Eqn.~(\ref{eq:gamma-turb}).\label{fig:fig9}}
\end{figure}

Figure~\ref{fig:fig9} shows the main result of the present study, which is the 
value of $U_{x,max}$ (defined in Eqn.~(\ref{eq:uxmax})) as a function of time in 
a lin-log plot, for runs corresponding to different turbulent intensities. The 
thick black lines corresponds to $U_{x,max}(t)$ for each simulation, the thin 
black lines indicate one standard deviation with respect to the average (i.e. 
$U_{x,max}\pm u_{turb}$), and the straight gray lines are the theoretical 
predictions for each case, as emerges from Eqn.~(\ref{eq:gamma-turb}). Note that 
the theoretical slopes (i.e. the gray lines in Figure~\ref{fig:fig9}) are not 
best fits to each of the simulations, but the result arising from 
Eqn.~(\ref{eq:gamma-turb}), which contains only one free parameter for the whole 
set of simulations, namely the constant $C$.  This constant is the only 
dimensionless parameter that remains undetermined by the dimensional analysis 
described above. We find that the value of $C$ that best fits all our 
simulations is $C\approx 18.8$.

\section{Discussion}\label{discu}

In the previous section, we presented results from numerical simulations showing 
the role of a background turbulence in reducing the growth rate of an ongoing KH 
instability. These numerical results are intended to simulate the KH instability 
being developed at the interface between some CMEs and the ambient corona, which 
have been recently reported in the literature. There is also mounting observational 
evidence about the turbulent nature of the solar corona, mostly related with spatially 
unresolved motions leading to measurable nonthermal broadenings in coronal spectral lines. 

To numerically model this turbulent background, we made a number of simplifying 
assumptions. For instance, we assume the turbulent regime to be spatially 
homogeneous and isotropic and also stationary. We maintain this turbulent state 
throughout the whole simulation by applying a stationary stirring force of 
intensity $f_{turb}$ at a well defined lengthscale $l_{turb}$. We deliberately 
chose this lengthscale to be much smaller than the wavelength of the KH unstable 
mode, since the AIA images reporting the KH pattern do not show any observable 
evidence of a turbulent background. Also, the rotation period of the 
energy-containing vortices is of the order of $\tau_{turb}\simeq 
l_{turb}/u_{turb}$, which remains shorter than the instability growth time for 
all the cases considered. The properties of this turbulent regime are therefore 
determined by only two input parameters: $l_{turb}$ which is kept fixed 
throughout the whole study, and $f_{turb}$ which is varied to give rise to cases 
with different turbulent velocities ($u_{turb}$) and effective viscosities 
($\nu_{turb}$). 

We can use Eqs.~(\ref{eq:uturb-f})-(\ref{eq:nuturb}) to express the effective 
viscosity $\nu_{turb}$ in terms of two measurable quantities such as $u_{turb}$ 
and $l_{turb}$. A crude estimate of the dimensionless constant in 
Eqn.~(\ref{eq:uturb-f}) leads to $u_{turb} \approx 0.22\ (f_{turb}\ 
l_{turb})^{1/2}$ and therefore
\begin{equation}\label{eq:nuturb-uturb}
 \nu_{turb} \approx 85.4\ u_{turb}\ l_{turb}\ .
\end{equation}

If we refer for instance to the KH event occurred on 2010 November 3 and reported by 
\citet{foullon2011}, they estimate a velocity jump at the interface of $U_0 = 
340\ km.s^{-1}$ and a wavelength for the KH pattern of $\lambda = 2\pi L_0 = 
18.5\ Mm$ (corresponding to a length unit of $L_0 = 3\ Mm$ and $k_y = 1$ in our 
simulations). For $k_y = 1$, the dispersion relation reduces to $\gamma \approx 
0.87 - \nu_{turb}$, as shown in Eqn.~(\ref{eq:gamma-turb}). The instability growth 
rate estimated by \citet{foullon2013} for this event is $\gamma \approx 0.033\ s^{-1}$, 
which in our dimensionless units becomes $\gamma L_0/U_0 = 0.29 = 0.87 - \nu_{turb}$.
From this expression we can estimate the value of $\nu_{turb}$ required to explain the 
growth rate observed for this particular KH event. More interestingly, using Eqn.~(\ref{eq:nuturb-uturb}) 
we can obtain a level of turbulent velocity of $u_{turb}\approx 47 km.s^{-1}$ (for the value of 
$l_{turb}$ used in our simulations), which is well within the range reported by \citet{doschek2014} 
from Hinode observations. It is important to recall that other effects besides turbulence 
might contribute to reduce the instability growth rate. Depending of the parameter values of the particular 
KH event being considered, the compressibility of the plasma or the strength of the magnetic field component 
along the shear flow might play a role.

Another consequence that we can derive from the present analysis is that, given the 
fact that the turbulence did not completely suppress the KH instability, we can in 
principle use Equations~(\ref{eq:gamma-turb})-(\ref{eq:nuturb-uturb}) to estimate an 
upper bound for $l_{turb}$ for any observed 
value of $u_{turb}$. For the turbulent attenuation to be negligible (i.e. $\nu_{turb}\ll 0.87$) 
and assuming a turbulent velocity of $60\ km.s^{-1}$ (see \citet{doschek2014}), we obtain for 
$l_{turb}$ an upper bound of $170\ km$. In general, 
\begin{equation}\label{eq:lturb}
 l_{turb}\ll 170\ km (\frac{u_{turb}}{60\ km.s^{-1}})^{-1}
\end{equation}
In summary, in order for the invoked turbulent state to produce nonthermal 
broadening of spectral lines of the order of $u_{turb}$ and at the same time not 
to affect the observed KH event in any appreciable manner, the typical size 
$l_{turb}$ of its energy-containing eddies should satisfy Eqn.~(\ref{eq:lturb}).

\section{Conclusions}\label{conclu}

The study presented in this paper was motivated by two relatively recent 
observational findings on the nature of the solar corona. One of them is the 
apparent development of the Kelvin-Helmholtz instability as some CMEs expand in 
the ambient corona, as shown by AIA/SDO images 
\citep{foullon2011,foullon2013,ofman2011}. The second one is that the coronal 
plasma seems to be in a turbulent state, as 
evidenced by the nonthermal broadening of coronal spectral lines measured from 
EIS/Hinode data \citep{doschek2008,brooks2011,tian2012,doschek2014}. 

Our main goal has been to study the feasibility for these two apparently 
dissimilar features to coexist. Namely, the large-scale laminar pattern observed 
for the KH instability, and the small-scale spatially unresolved turbulent 
motions leading to the observed nonthermal broadenings. We therefore performed 
three-dimensional simulations of the MHD equations, to study the evolution of 
the KH instability in the presence of a turbulent ambient background for 
different intensities of this turbulence.

Theoretically, the effect of a small-scale turbulence on a large-scale flow 
would be to produce an enhanced diffusivity which can be modeled by an effective 
or turbulent viscosity. The impact of this small-scale turbulence on an ongoing 
large-scale instability such as KH, would then be a reduction of its growth 
rate, as emerges from Eqn.~(\ref{eq:gamma-turb}). The degree of this reduction 
is controlled by the turbulent viscosity $\nu_{turb}$ which we obtained from a 
dimensional analysis to be $\nu_{turb} = C (f_{turb}\ l_{turb}^3 )^{1/2}$ (see 
Eqn.~(\ref{eq:nuturb})), leaving only the dimensionless constant $C$ 
undetermined. 

The comparison between the instability growth rates obtained from our 
simulations with the ones arising from Eqn.~(\ref{eq:gamma-turb}) esentially 
confirms this theoretical scenario, while providing an empirical determination 
for the dimensionless constant $C$, which amounts to $C\approx 18.8$. Perhaps 
more importantly, since $\nu_{turb} \propto u_{turb}\ l_{turb}$ and given the 
fact that the instability has not been completely quenched by the turbulence 
(otherwise it would not have been observed), observational determinations of 
$u_{turb}$ from nonthermal broadenings pose an upper limit to the correlation 
length of the turbulence $l_{turb}$. For observational values of $u_{turb} 
\approx 20-60\ km.s^{-1}$, the correlation length of turbulence is expected to 
be smaller than about $l_{turb}\approx 170-510\ km$, which is consistent with not 
having been spatially resolved by current coronal imaging spectrometers such as 
EIS aboard Hinode.	

\acknowledgments
DG and EED acknowledge financial support from grant SP02H1701R from 
Lockheed-Martin to SAO. DG also acknowledges support from PICT grant 0454/2011 
from ANPCyT to IAFE and PDM acknowledges support from PICTs 2011-1529 and 
2011-1626 from ANPCyT to IFIBA (Argentina).

\clearpage
\begin{deluxetable}{lccccccc}
 \tabletypesize{\scriptsize}
\tablecaption{Values of dimensionless parameters for the simulations: $N$ is the 
linear size, $U_0$ is the velocity at each side of the shear layer, $\Delta$ is 
the thickness of the shear layer, $v_A^\parallel$ is the parallel component of the Alfv\'en speed, $\eta$ is the 
magnetic diffusivity, $\nu$ is the kinematic viscosity, $l_{turb}$ is the the 
length scale of the turbulence and $f_{turb}$ is the strength of the turbulent 
forcing.\label{table}}
\tablewidth{0pt}
\tablehead{
\colhead{N} & \colhead{$U_0$} & \colhead{$\Delta$} & \colhead{$v_A^\parallel$} & 
\colhead{$\eta$} &
\colhead{$\nu$} & \colhead{$l_{turb}$} & \colhead{$f_{turb}$} 
}
\startdata
256 & 1 & 0.1 & 0.2 & $2.10^{-3}$ & $2.10^{-3}$ & 0.05 & 0, 2, 3, 4, 5, 10, 15 
\\
\enddata
\end{deluxetable}


\begin{thebibliography}{}
\bibitem[Balbus \& Hawley(1998)]{balbus1998}Balbus, S.A., \& Hawley, J.F. 1998, 
\rmp, 70, 1
\bibitem[Bartoe(1982)]{bartoe1982}Bartoe, J.D. 1982, \asr, 2, 185
\bibitem[Begelman, Blandford \& Rees(1984)]{begelman1984}Begelman, M.C., 
Blandford, R.D., \& Rees, M.J. 1984, \rmp, 56, 255
\bibitem[Bodo et al.(1994)]{bodo1994}Bodo, G., Massaglia, S., Ferrari, A., \& 
Trussoni, E. 1994, \aap, 283, 655
\bibitem[Braginskii(1965)]{braginskii1965}Braginskii, S.I. 1965, \rpp, 1, 205
\bibitem[Brandt \& Mendis(1979)]{brandt1979}Brandt, J.C., \& Mendis, D.A. 1979, 
``Solar System Plasma Physics'', Eds. C.F. Kennel et al. (North Holland: 
Amsterdam), 2, 253
\bibitem[Brooks and Warren(2011)]{brooks2011} Brooks, D.H., \& Warren, H.P. 
2011, \apjl, 727, L13
\bibitem[Chandrasekhar(1961)]{chandra1961}Chandrasekhar, S. 1961, ``Hydrodynamic 
and Hydromagnetic Stability'', Oxford University Press: New York.
\bibitem[Coyner \& Davila(2011)]{coyner2011}Coyner, A.J., \& Davila, J.M. 2011, 
 \apj , 742, 115
\bibitem[Doschek et al.(2008)]{doschek2008} Doschek, G.A., Warren, H.P., 
Mariska, J.T., Muglach, K., Culhane, J.L., Hara, H., \& Watanabe, T. 2008, \apj, 
686, 1362
\bibitem[Doschek et al.(2014)]{doschek2014} Doschek, G.A., McKenzie, D.E., \& 
Warren, H.P. 2014, \apj, 788, 26
\bibitem[Drazin(1958)]{drazin1958}Drazin, P.G. 1958, \jfm, 4, 214
\bibitem[Drazin \& Reid(1981)]{drazin1981}Drazin, P.G., \& Reid, W.H. 1981, 
``Hydrodynamic Stability'', Cambridge University Press: Cambridge.
\bibitem[Dwarkadas \& Balbus(1996)]{dwarkadas1996}Dwarkadas, V.V., \& Balbus, 
S.A. 1996, \apj, 467, 87
\bibitem[Ershkovich, Nusinov \& Chernikov(1973)]{ershkovich1973}Ershkovich, A.I., Nusinov, A.A., \& 
Chernikov, A.A. 1973, \sa, 16, 705
\bibitem[Fairfield et al.(2000)]{fairfield2000}Fairfield, D.H., Otto, A., Mukai, 
T., Kokubun, S., Lepping, R.P., Steinberg, J.T., Lazarus, A.J., Yamamoto, T. 
2000, \jgr, 105, 21159
\bibitem[Ferrari, Trussoni \& Zaninetti(1980)]{ferrari1980} Ferrari, A., 
Trussoni, E., \& Zaninetti, L. 1980, \mnras, 193, 469
\bibitem[Foullon et al.(2011)]{foullon2011} Foullon, C., Verwichte, 
E.,Nakariakov, V.M., Nykyri, K., \& Farrugia, C.J. 2011, \apjl, 729, L8
\bibitem[Foullon et al.(2013)]{foullon2013} Foullon, C., Verwichte, E., Nykyri, 
K., Aschwanden, M.J., \& Hannah, I.G. 2013, \apj, 767, 170
\bibitem[Fujimoto \& Teresawa(1995)]{fujimoto1995}Fujimoto, M., \& Teresawa, T. 
1995, \jgr, 100, 12025 
\bibitem[Gonzalez \& Gratton(1994)]{gonzalez1994} Gonzalez, A.G., \& Gratton, J. 
1994, \jpp, 52, 223
\bibitem[G\'omez et al.(2005)]{gomez2005} G\'omez, D.O., Mininni, P.D., \& 
Dmitruk, P. 2005, Phys. Scripta, T116, 123
\bibitem[G\'omez et al.(2014)]{gomez2014} G\'omez, D.O., Bejarano, C., \& 
Mininni, P.D. 2014, Phys. Rev. E, 89, 053105
\bibitem[Hasegawa(1985)]{hasegawa1985}Hasegawa, A. 1985, \ap, 34, 1
\bibitem[Helmholtz(1868)]{helmholtz1868} Helmholtz, H.L.F. 1868, Monthly Rep. 
Royal Prussian Acad. Phil. Berlin, 23, 215
\bibitem[Lau \& Liu(1980)]{lau1980}Lau, Y.Y., \& Liu, C.S. 1980, \pof, 23, 939
\bibitem[Mariska(1992)]{mariska1992}Mariska, J.T. 1992, in {\it The Solar Transition 
Region} (Cambridge: Cambridge Univ. Press).
\bibitem[Masters et al.(2010)]{masters2010}Masters, A., and 10 co-authors 2010, 
\jgr, 115, A07225
\bibitem[Miura \& Pritchett(1982)]{miura1982}Miura, A., \& Pritchett, P.L. 1982, 
\jgr, 87, 7431
\bibitem[Miura(1992)]{miura1992}Miura, A. 1992, \jgr, 97, 10655
\bibitem[Mostl et al.(2013)]{mostl2013}Mostl, U.V., Temmer, M., \& Veronig, A.M. 
2013, \apj, 766, L12
\bibitem[Nykyri \& Otto(2001)]{nykyri2001}Nykyri, K., \& Otto, A. 2001, \grl, 
28, 3565
\bibitem[Nykyri \& Foullon(2013)]{nykyri2013}Nykyri, K., \& Foullon, C. 2013, \grl, 
40, 4154
\bibitem[Ofman \& Thompson(2011)]{ofman2011}Ofman, L., \& Thompson, B.J. 2011, 
\apjl, 734, L11
\bibitem[Parker(1958)]{parker1958}Parker, E.N. 1958, \apj, 128, 6640
\bibitem[Poedts, Rogava \& Mahajan(1998)]{poedts1998}Poedts, S., Rogava, A.D., 
\& Mahajan, S.M. 1998, \apj, 505, 369
\bibitem[Prialnik et al.(1986)]{prialnik1986}Prialnik, D., Eviatar, A., \& Ershcovich, 
A.I. 1986, \jpp, 35, 209
\bibitem[Sundberg et al.(2011)]{sundberg2011}Sundberg, T., and 7 co-authors 
2011, \pss, 59, 2051 
\bibitem[Teriaca et al.(1999)]{teriaca1999}Teriaca, L., Doyle, J.G., Erdelyi, R., \& 
Sarro, L.M. 1999, \aanda, 352, L99
\bibitem[Tian et al.(2012)]{tian2012} Tian, H., McIntosh, S.W., Xia, L., He, J., 
\& Wang. X. 2012, \apj, 748, 106
\bibitem[Kelvin(1871)]{kelvin1871} Thomson, W.(Lord Kelvin) 1871, Phil. Mag., 
42, 362
\bibitem[Wyper \& Pontin(2013)]{wyper2013} Wyper, P.F., \& Pontin, D.I. 2013, \pop, 
20, 032117
\end{thebibliography}
\end{document}